\documentclass[aps,prl,twocolumn,showpacs,superscriptaddress]{revtex4}
\usepackage{amssymb}
\usepackage{graphicx}
\usepackage{amsmath}
\usepackage{textcomp}
\usepackage{lscape}

\newcommand{\bgreek}[1]{\mbox{\boldmath$#1$\unboldmath}}

\begin{document}
\title{Theory of the spin relaxation of conduction electrons in silicon}

\author{J. L. Cheng}
\affiliation{Hefei National Laboratory for Physical Sciences at
  Microscale and Department of Physics, University of Science
  and Technology of China, Hefei,  Anhui, 230026, China}
\affiliation{Institute for Theoretical Physics, University of
  Regensburg, 93040 Regensburg, Germany}
\author{M. W. Wu}%
\affiliation{Hefei National Laboratory for Physical Sciences at
  Microscale and Department of Physics, University of Science
  and Technology of China, Hefei,  Anhui, 230026, China}
\author{J. Fabian}
\thanks{Author to whom correspondence should be addressed}%
\email{jaroslav.fabian@physik.uni-regensburg.de}
\affiliation{Institute for Theoretical Physics, University of
  Regensburg, 93040 Regensburg, Germany}

\begin{abstract}
A realistic pseudopotential model is introduced to investigate the
phonon-induced spin relaxation of conduction electrons in bulk silicon.
We find a surprisingly subtle interference of the Elliott and Yafet processes
affecting the spin relaxation over a wide temperature range, suppressing
the significance of the intravalley spin-flip scattering, previously considered
dominant, above roughly 120 K.
The calculated spin relaxation times $T_1$ agree with the spin resonance and spin
injection data, following a $T^{-3}$ temperature dependence. The valley anisotropy
of $T_1$ and the spin relaxation rates for hot electrons are predicted.
\end{abstract}


\maketitle

Silicon is the core material for the information technology. Yet we know
surprisingly little about the spin relaxation processes of its conduction electrons
\cite{zutic04, fabian07}. With the pioneering demonstration of the spin injection
into silicon by Appelbaum et al. \cite{appelbaum, zutic07} and related experimental
breakthroughs \cite{jonker, huang} and steps towards silicon spintronics \cite{jansen}, as
well as with theoretical analyses \cite{zutic06, mavropoulos,zhang},
we have gained valuable new insight into the spin transport and relaxation in this material.

A systematic investigation of the conduction electron spin relaxation
time $T_1$ in silicon was conducted earlier by L\'epine  \cite{lepine}
using electron spin resonance (ESR)
(see the summary in Ref. \onlinecite{fabian07}). As the ESR does not discriminate
between conduction electrons and electrons bound on donors, the representative
data are limited to temperatures above $T \approx 150$ K at which most electrons are
in the conduction band for the investigated donor densities \cite{lepine}. Spin injection,
on the other hand, looks at conduction electrons only.
Appelbaum et al. \cite{appelbaum,huang} extracted useful data below $150$ K, filling the gap. These measurements were using samples with nondegenerate electron densities. The conduction electron spin relaxation and
spin transport properties were also investigated in Si/SiGe quantum wells \cite{tyryshkin,zhang}.
There, however, the spin coherence is due not to the bulk-derived properties, but
rather due to the appearance of the structure-inversion anisotropy spin-orbit fields.

While it is generally believed that the spin relaxation in silicon is caused by the
Elliott-Yafet mechanism \cite{lepine,fabian07,yafet,appelbaum,huang} (unlike in
III-V semiconductors, in which the most important mechanism is the
D'yakonov-Perel' one \cite{jiang})  there is yet
no systematic theoretical study of it. There are two processes involved:
that of Elliott and Yafet. In the Elliott processes \cite{elliott}
the spin-flip is due to the admixture of the Pauli up and down spins in the
Bloch state, caused by spin-orbit coupling. The electron-phonon matrix element
couples only equal Pauli spins. In the Yafet processes \cite{yafet} spin flips are
due to the phonon-modulated spin-orbit coupling so that the electron-phonon coupling
alone couples opposite spins. The two processes interfere destructively at low phonon
momenta affecting $T_1$ typically at very low temperatures \cite{yafet}.
Yafet gave qualitative estimates for $T_1$ in silicon assuming intravalley electron-acoustic phonon
scattering, finding that $T_1 \sim T^{-5/2}$. This temperature dependence has been widely
used to fit experimental data \cite{lepine,huang,fabian07,appelbaum,lancaster}.

Here we perform comprehensive theoretical investigation of $T_1$ in bulk silicon within
the Elliott-Yafet mechanism. We introduce a pseudopotential model that reproduces the known spin-orbit
splittings of the relevant electronic states. The model, together with a realistic
phonon structure taken from the adiabatic bond charge model \cite{weber}, allows us
to calculate the spin mixing probabilities and the electron-phonon-induced spin flips for both the Elliott and Yafet processes. We show that the interference between the two processes
affects $T_1$ over a remarkably wide temperature range. Both the intra and intervalley spin-flip scatterings are important [optical (OP) phonons  are less relevant than acoustic (AC) ones] making Yafet's prediction invalid. Our calculated $T_1(T)$, which is in quantitative agreement with experiment,
is well described by the  $T_1 \sim T^{-3}$ dependence. We further predict the valley anisotropy
of $T_1$ and give the spin relaxation rates for hot electrons.

The presence of space inversion symmetry in bulk silicon allows to write
the Bloch states as combinations of the Pauli spinors \cite{yafet, fabian07},
\begin{eqnarray}
  |\mathbf k,n\Uparrow\rangle &=& \sum_{\mathbf g}\big[a_{\mathbf
    k,n}(\mathbf g)|\uparrow\rangle + b_{\mathbf
    k,n}(\mathbf g)|\downarrow\rangle\big]|\mathbf k+\mathbf g\rangle\ ,\\
  |\mathbf k, n\Downarrow\rangle &=& \sum_{\mathbf g}\big[a^{\ast}_{\mathbf
    k,n}(\mathbf g)|\downarrow\rangle - b^{\ast}_{\mathbf
    k,n}(\mathbf g)|\uparrow\rangle\big]|\mathbf k+\mathbf g\rangle\ .
\end{eqnarray}
Here $\mathbf k$ is the lattice momentum confined to the first Brillouin
Zone, $n$ is the band index, $\Uparrow$($\Downarrow$) is the effective spin
index, $\mathbf g$ denote the reciprocal lattice vectors, and $|\mathbf
k\rangle$ stand for the plane waves. The two states above are
degenerate, and can so be chosen as the spin ``up''
and ``down'' states to satisfy  $\langle\mathbf
k,n\Uparrow|\sigma_z|\mathbf k,n\Downarrow\rangle = 0$. The mixing
of the Pauli spins in the spin ``up'' (``down'') state is characterized
by the mixing probability $|b_{\mathbf   k,n}|^2$, which is key to understand spin relaxation.

We follow the scheme of Ref. \cite{fabian99} and build a pseudopotential
model incorporating spin-orbit coupling to obtain the electronic states
and electron-phonon coupling needed to calculate $T_1$.
The pseudopotential of each atom is the sum of the scalar ($v_{n}$)
and spin-orbit ($v_{\rm so}$) parts, $v(\mathbf r)=v_n(\mathbf r) +
v_{\rm so}(\mathbf r)$; the form factors are $v(\mathbf
k_1,\mathbf k_2)=\int (d^3 \mathbf r /a^3)
e^{-i(\mathbf k_1 - \mathbf k_2) \cdot\mathbf r}v(\mathbf r)$,
with the lattice constant $a=5.431$ $\AA$.
The scalar part $v_n(\mathbf r)$ is taken from
Ref.\ [\onlinecite{chelikowsky}], reproducing well the silicon bands.
We introduce the spin-orbit part as \cite{fabian98,weisz}
$v_{\rm so}(\mathbf r)=\lambda \theta(r-r_c)\mathbf{\hat{L}}\cdot\hat{\bgreek{\sigma}}{\cal
  P}_1$, where $\mathbf{\hat{L}}$ is the angular momentum
operator, ${\cal P}_l$ is the projector on the orbital
momentum state $l$, $r_c=2r_B$ is the effective radius for the spin-orbit
coupling ($r_B$ is the Bohr radius), and $\lambda$ is a spin-orbit parameter
to be fitted to the valence band spin-orbit splitting $\Delta_{25}^{l}$.
The fitted form factor of $v_{\rm so}$ can be approximated by \cite{weisz} $v_{\rm so}
(\mathbf k_1,\mathbf k_2)=-i\Delta_{\rm so} (a/2\pi)^2\mathbf k_1\times\mathbf
k_2\cdot\bgreek{\sigma}$, with the effective spin-orbit interaction
$\Delta_{\rm so}=(16\pi^3/45) (r_c/a)^5 \lambda=9.475\times10^{-4}$\ eV.
The actual spin-orbit pseudopotential Hamiltonian matrix element comprises
the structure and form factors:
$\langle \mathbf k_1|H_{\rm so}|\mathbf k_2\rangle=\cos[(\mathbf k_1-\mathbf
k_2)\cdot\bgreek{\tau}]v_{\rm so}(\mathbf k_1,\mathbf k_2)$, where
$\bgreek{\tau}= (a/8)(1,1,1)$.
The pseudopotential Hamiltonian is diagonalized on a basis of 387 plane-waves, large enough to converge the spin mixing probabilities around the conduction band edge. The sums over
momenta are performed using the tetrahedron method.

\begin{figure}[htp]
  \centering
  \includegraphics[height=8cm,angle=-90]{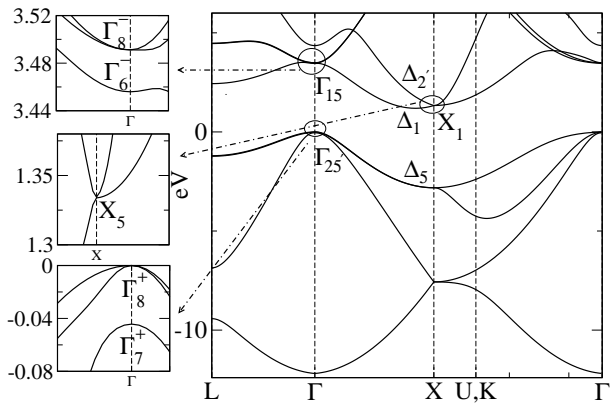}
  \caption{Calculated electronic band structure of silicon. The single group
    notation \cite{cardona} is used in the main plot, while the double group
    notation \cite{peteryu} appears on the zoom plots.}
  \label{fig:band}
\end{figure}

\begin{table}[h]
  \centering
  \begin{tabular}[t]{c|ccc|cc|cc|c}
    \hline
    Unit&\multicolumn{3}{c}{eV}\vline&\multicolumn{2}{c}{meV}\vline&\multicolumn{2}{c}{$m_0$}\vline&$\Gamma X$\\
    \hline
    &$E_{\Gamma_{15}}$&$E_{X_1}$&$E_{g}$&$\Delta_{25}^l$&$\Delta_{15}$&$m_{l}^{\Delta}$&$m_{t}^{\Delta}$&$ k_{\mathtt{min}}$\\
    \hline
    Exp. \cite{landolt}&3.4&1.25&1.17&44&30-40&0.9163&0.1905&\\
    Cal.&3.46&1.33&1.19&44&35&0.915&0.204&0.846\\
    \hline
  \end{tabular}
  \caption{Comparison between measured and calculated band-structure
  characteristics. Displayed is the energy
  $E_{\Gamma_{15}}$ of the direct gap at $\Gamma_{15}$, the energy
  $E_{X_1}$ of the $X_1$ ($X_5$) point, and the indirect band gap $E_g$.
  The spin-orbit split off energy of the top of the valence band is $\Delta^l_{25}$,
  and that of the conduction band $\Gamma_{15}$ point is $\Delta_{15}$ (the stated
  range of values are calculated \cite{landolt}). Further
  shown are the longitudinal ($m_l^{\Delta}$) and transverse ($m_t^{\Delta}$)
  effective masses (in the units of the free electron mass) at the conduction
  band minimum at $k_{\rm min}$ along the $\Delta$ lines $\Gamma X$. }
  \label{tab:band}
\end{table}

The calculated electronic band structure is shown in
Fig. \ref{fig:band};  Table \ref{tab:band} displays
selected band-structure properties, calculated and measured.
The agreement with known data is very satisfactory, justifying our
pseudopotential for exploring spin-orbit
effects in silicon. In the following we denote the wave vectors of
the six conduction-band valleys as $\mathbf K_i$,
with $i=X(\bar{X})/Y(\bar{Y})/Z(\bar{Z})$ standing for the corresponding
valley orientation.

\begin{figure}[htp]
  \centering
  \includegraphics[height=6cm,angle=0]{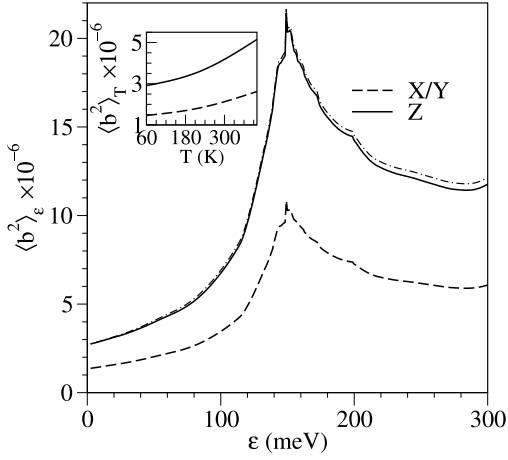}
  \caption{Calculated energy dependence of the spin mixing probability
  $\langle b^2\rangle_\epsilon$ in the $X/Y$ (dashed) and $Z$ (solid) valleys. The
    dot-dashed curve is $2\langle b^2\rangle_{\epsilon,X}$. The inset
    shows the temperature dependent $\langle b^2\rangle_T$.}
  \label{fig:abk2}
\end{figure}

Let us first see what can we learn about $b^2$ from symmetry group
arguments. Consider the $Z$ valley and take the spin-orbit interaction
$H_{\mathtt{so}}\propto L_x\sigma_x+L_y\sigma_y+L_z\sigma_z$ as a
perturbation. Group theory shows that the conduction band
($\Delta_1$, see Fig.~\ref{fig:band}) couples only to the
the valence band ($\Delta_5$) at the band
edge, so that the spin mixing is of the order
of $|b_{ck}|^2\approx (\Delta_{\rm so}/E_g)^2\sim10^{-6}$
\cite{elliott,yafet}. In addition to the magnitude, we can also
learn about the valley anisotropy of $b^2$. We can describe
the symmetry character of the conduction band orbital states
as $|Z\rangle$, and that of the degenerate valence states as $|X\rangle$/$|Y\rangle$ \cite{peteryu},
in the usual ${\bf k} \cdot {\bf p}$ theory sense.  The only
nonvanishing matrix elements of the orbital momentum between these
states are $\langle X|L_y|Z\rangle = -\langle Y|L_x|Z\rangle$, by symmetry.
Therefore, the effective spin-orbit interaction involves only
the $\sigma_{x}$ and $\sigma_y$ terms with equal contribution. If the spin is quantized along $z$, both $\sigma_x$ and $\sigma_y$
terms contribute equally to the spin mixing $b^2$. However, only one of them,
{\it i.e.,} $\sigma_y$/$\sigma_x$ term, contributes to the spin
mixing for the $x$/$y$ quantized spin.
Since the spin along $x$/$y$ for the $Z$ valley is
equivalent to the spin along $z$ for the $X$/$Y$ valley under
a $\frac{\pi}{2}$-rotation around the $y$/$x$ axis, we can
conclude that the average spin mixing is
anisotropic with respect to the valley orientation: $\langle
b^2\rangle_{\epsilon,Z}\approx2\langle b^2\rangle_{\epsilon,X/Y}$.
This is indeed found from numerics, as shown in Fig.~\ref{fig:abk2}.
See EPAPS Document No. for a more intuitive explanation of the 
anisotropy.

Figure ~\ref{fig:abk2} also shows a peak around $150$~meV, which is roughly the energy
of $X_1$. This peak indicates a large spin mixing on the $X$ plane.  In the $X$ plane, there is
anticrossing along the X-W direction due to spin-orbit coupling of
degenerate bands $\Delta_1$ and $\Delta_2^{\prime}$, which results in
large spin mixing (similar to spin hot spots \cite{fabian98}). The
anticrossing is absent at the X-point, but grows along the X-W
direction  to the order of $\frac{1}{2}$ quickly. See EPAPS Document
No. for more details on spin hot spots  in silicon.
In the inset of Fig. \ref{fig:abk2}, we also plot the temperature dependence of
the average spin mixing $\langle b^2\rangle_{T,i}=\sum_{\mathbf k\in i}|b_{c\mathbf
  k}|^2f(\varepsilon_\mathbf k)/\sum_{\mathbf k\in i}f(\varepsilon_\mathbf k)$, where $f(\varepsilon_{\mathbf k}) = C  e^{\varepsilon_{\mathbf k}/(k_BT)}$ is the properly normalized ($C$) Maxwell-Boltzmann distribution.
  The anisotropy, $\langle b^2\rangle_{T,Z}\approx 2\langle b^2\rangle_{T,X/Y}$,
  remains.

The spin-flip electron-phonon matrix elements for the conduction states $|\mathbf k_1;c\Uparrow\rangle$ and $|\mathbf k_2;c\Downarrow\rangle$ are \cite{yafet}
\begin{eqnarray}
&&M^{\lambda}(\mathbf k_1,\mathbf
k_2)=  -i\sqrt{\frac{\hbar}{\rho\Omega^{\lambda}_{\mathbf q}}}\sum_{\mathbf
  g_1\mathbf g_2}\bigg[\Delta\mathbf
k\cdot\sum_{\alpha}\bgreek{\xi}_{\mathbf q,\alpha}^{\lambda}e^{-i\Delta\mathbf
      k\cdot\bgreek{\tau}_{\alpha}}\bigg]\nonumber\\
    &&\times \begin{pmatrix}a_{\mathbf k_1,c}(\mathbf g_1) \\
  b_{\mathbf k_1,c}(\mathbf g_2)\end{pmatrix}^{\dag}v(\mathbf
k_1+\mathbf g_1,\mathbf k_2+\mathbf g_2)\begin{pmatrix}-b_{\mathbf
    k_2,c}^{\ast}(\mathbf g_2)\\a_{\mathbf k_2,c}^{\ast}(\mathbf g_2) \end{pmatrix},
\label{eq:scatm}
\end{eqnarray}
where $\mathbf q=\mathbf k_1-\mathbf k_2$ is the phonon wave vector,
$\bgreek{\tau}_\alpha=\pm\bgreek{\tau}$ are the position vectors of
the two basis atoms, $\Delta\mathbf k=\mathbf
k_1+\mathbf g_1-\mathbf k_2-\mathbf g_2$, $\rho$ is the silicon
density, and  $\hbar\Omega_{\mathbf
  q}^{\lambda}$ and $\bgreek{\xi}_{\mathbf
  q,\alpha}^{\lambda}$ are the phonon energy and the polarization vector
for the $\alpha$th atom, obtained here from the adiabatic bond-charge model \cite{weber}.
For the Elliott processes  $v = v_n$, while for Yafet processes
$v = v_{\rm so}$ in Eq. (\ref{eq:scatm}).

The spin relaxation rate is given by
$T_1^{-1}(T)=\sum_{i,j}T_{1,ij}^{-1}$, with
\begin{equation}
  T_{1,ij}^{-1}(T)=\sum_{\lambda}\int
  d\epsilon\Gamma^{\lambda}_{ij}(\epsilon, T) f(\epsilon).
\end{equation}
Here $\Gamma^{\lambda}_{ij}$ is the
total scattering rate for
the electrons of energy $\epsilon$ making $\lambda$-phonon assisted
spin-flip transitions from the $i$th to the  $j$th valley.
For non-degenerate electron densities,
\begin{eqnarray}
  \label{eq:fermirule}
  \Gamma^{\lambda}_{ij}(\epsilon, T)&=&\frac{4\pi}{\hbar}\sum_{\mathbf
    k_1\in i}\delta(\varepsilon_{c\mathbf k_1}-\epsilon)\sum_{\mathbf
    k_2\in j}\sum_{\pm}|M^{\lambda}(\mathbf k_1,\mathbf k_2)|^2\nonumber\\
  &\times&({\bar n}_{{\mathbf  q}\lambda} +
  \frac{1}{2}\pm\frac{1}{2})\delta(\varepsilon_{c\mathbf
    k_2}-\epsilon\pm\Omega_{{\mathbf q}\lambda}),
\end{eqnarray}
where $\bar{n}_{\mathbf q,\lambda}$ is the phonon
distribution. Considering the low-energy intravalley processes with AC phonons
Yafet found the dependence of $T_1^{\rm intra} \sim T^{-5/2}$ \cite{yafet}.
That prediction was based on Yafet's observation that the time reversal and space
inversion symmetry inhibits the low momentum scattering, expressed
in the quadratic dependence, $M^{\lambda}(\mathbf k_1,\mathbf
k_2)\propto {|\mathbf q|^2}/\sqrt{\Omega_{\mathbf q}^{\lambda}}$,
valid for low $\mathbf q$.

\begin{figure}[htp]
  \centering
  \includegraphics[height=6.5cm]{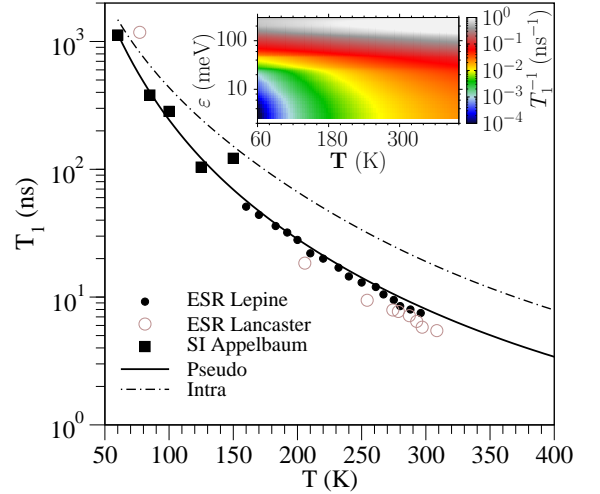}
  \caption{Spin relaxation time in silicon as a function of temperature. The
  solid curve is the calculation, the symbols are the spin injection (Appelbaum)
  \cite{appelbaum, corrected} and the electron spin resonance (Lepine and Lancaster) \cite{lepine,lancaster} data
  (see Ref. \onlinecite{fabian07}). The dashed-dotted curve is $T_1^{\mathtt{intra}}$ of the intravalley scattering only. The inset
  shows the contour plot of the spin relaxation rate $T_1^{-1}$ of
  hot electrons, as
 a function of the electron energy $\varepsilon$ and lattice
   temperature $T$. }
  \label{fig:srt}
\end{figure}

Our main result is in Fig. \ref{fig:srt} which shows the calculated $T_1(T)$.
The agreement with experiments \cite{lepine,fabian07,lancaster,appelbaum} is
very good. Clearly, the intravalley spin relaxation $T_1^{\mathtt{intra}}$ alone is
not sufficient to explain the experiment. While there is no single scattering type
governing the whole temperature
range, precluding a simple theoretical prediction for $T_1(T)$, a
fit to the numerical data gives $T_1\sim T^{-3}$ in the investigated
temperature region. Considering intravalley processes only,
Yafet's prediction $T_1^{\mathtt{intra}}\propto T^{-5/2}$ works at low $T$.
Due to the strong spectral dependence (peak) of $\langle b^2\rangle_{\epsilon}$ more energetic
electrons at higher temperatures strongly bias the average spin relaxation, making
$T_1^{\mathtt{intra}}(T)$ decay faster than predicted. This behavior is also
reflected in the energy-resolved spin relaxation rate, shown in the inset
to Fig. \ref{fig:srt}. The strong increase of the rate with increasing energy is due
to the increase in the scattering phase space and $b^2$.

\begin{figure}[htp]
  \centering
  \includegraphics[height=7cm,angle=-90]{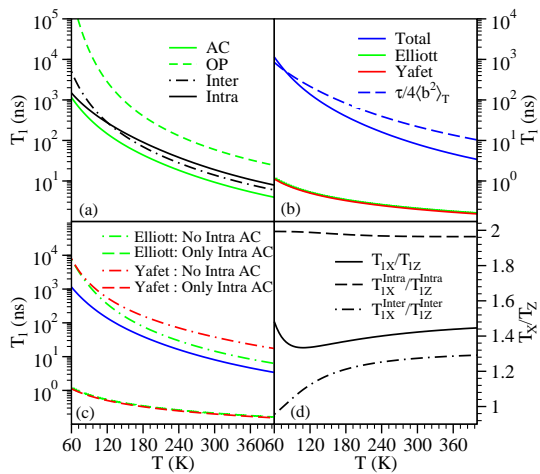}
  \caption{Analysis of the spin relaxation in silicon. (a) Spin relaxation
  time induced by acoustic (green solid) and  optical (green dashed) phonons,
  as well as by intervalley (dash-dotted) and intravalley (solid) scattering.
  (b) Elliott (green) and Yafet (red) processes compared to the total $T_1$ (blue).
  The estimate based on the knowledge of the momentum relaxation time $\tau$ and $b^2$ is also
  indicated (dashed blue). (c) Elliott and Yafet processes with and without intravalley acoustic phonons. The blue curve is the total $T_1$.
  (d) Valley-anisotropy  ratios $T_{1X}/T_{1Z}$ (solid); intravalley (dashed) and intervalley (dot-dashed) scattering are distinguished. }
  \label{fig:detail}
\end{figure}

Figure \ref{fig:detail} gives a comprehensive analysis of the calculated $T_1$.
The comparison of the AC and OP phonon contributions as well as the intra and intervalley
  contributions to
spin relaxation is shown in Fig. \ref{fig:detail}(a). We find that the spin
relaxation is dominated by intravalley electron-AC phonon scattering at
low temperatures ($T \alt 120$ K), and by intervalley electron-AC phonon
scattering at high temperatures ($T \agt 120$\ K). Figure \ref{fig:detail}(b)
resolves the individual contributions of the Elliott and Yafet processes.
Remarkably, they would individually lead to $T_1$ orders of magnitude below
the actual spin lifetime, over the whole temperature range! The destructive interference
of the two processes is a very subtle effect (also in terms of numerics).
The fact that the Elliott and Yafet processes interfere destructively
was predicted already by Yafet on symmetry grounds \cite{yafet}. See EPAPS Document
No. for more details.
The same figure also brings the estimated spin relaxation time
using the conventional formula \cite{fabian98,fabian99}
$T_1 = \tau /4\langle b^2\rangle_T$, where $\tau$ is the momentum relaxation
time calculated from our pseudopotential model.
This estimate fails in silicon especially at high temperatures.
The subtle nature of the interference between the Elliott and Yafet processes
is displayed in Fig. \ref{fig:detail}(c). Each process is dominated by the
the intravalley electron-AC phonon scattering over all temperatures.
The interference drastically reduces the significance of this scattering.
One consequence of our calculation would be the decrease of the Elliott-Yafet $T_1$
in silicon structures with reduced spatial inversion symmetry (gated, for
example), since in such cases the interference could be gradually removed.
A recent experiment indeed finds a dramatic reduction of the
spin relaxation time in Si/SiO$_2$ interfaces compared to the bulk \cite{Jang2009}.

Finally, we explore the anisotropy of $T_1$ with respect to the valley orientation.
Figure \ref{fig:detail}(d) gives the ratio $T_{1X}/T_{1Z}$ resolved for the intra and intervalley scattering. Assuming that the anisotropy of $T_1$ comes mainly from the spin mixing,
that is, $T_{1,ij}^{-1}\propto\sqrt{\langle b^2\rangle_{T,i}\langle
b^2\rangle_{T,j}} $, we get $T_{1X}^{\mathtt{intra}}/T_{1Z}^{\mathtt{intra}}\approx 2$, as
well as $T_{1X}^{\mathtt{inter}}/T_{1Z}^{\mathtt{inter}}\approx 4-2\sqrt{2}$
for intervalley $f$-processes (non-opposite valleys), and
$T_{1X}^{\mathtt{inter}}/T_{1Z}^{\mathtt{inter}}\approx 2$ for $g$-processes (opposite valleys). These estimates agree with our numerical results. The predicted anisotropy could
be tested in strained silicon with lifted valley degeneracy.

In summary, we explained the measured spin relaxation
in silicon performing realistic pseudopotential calculations of the Elliott-Yafet
mechanism. We found the dominant scattering processes and their temperature
 dependence, predicted the valley anisotropy as well as $T_1$ for hot electrons.

This work was supported by DFG SPP1285, GRK 638, the Natural Science Foundation of China
under Grant No. 10725417, the National Basic Research Program of China
under Grant No. 2006CB922005, and the Knowledge Innovation Project of
the Chinese Academy of Sciences.  J.L.C was partially supported by
China Postdoctoral Science Foundation.  We thank C. Ertler and
M. Q. Weng for valuable discussions.

\end{document}